\documentclass[aps,preprintnumbers,amsmath,amssymb]{revtex4}
\def\be{\begin{equation}}
\def\ee{\end{equation}}
\def\beq{\begin{eqnarray}}
\def\eeq{\end{eqnarray}}

\def\bay{\begin{array}}
\def\eay{\end{array}}

\newcommand{\cat}[1]{\mathbb{#1}}

\newcommand{\funct}[1]{\mathfrak{#1}}
\newcommand{\funcat}[2]{\cat{#2}^{\cat{#1}}}
\newcommand{\id}{\mathrm{id}}

\newtheorem{defin}{Definition}

\newcommand{\morph}[1]{\xrightarrow{#1}}
\newcommand{\morphr}[1]{\xrightarrow{\;#1\;}}

\newcommand{\dmorphr}[2]{{{{\raisebox{-0.28ex}{$\mathrm{#1}$}\atop
{\raisebox{-0.8ex}{$\rightrightarrows$}}}}\atop{\raisebox{-0.28ex}{$\mathrm{#2}$}}}}

\begin{document} %\openup8pt %This produces double-spaced document
%%%%%%%%%%%%%%%%%%%%%%%%%%%%%%%%%%%%%%%%%%%%%%%%%%%%%%%%%%%%%%%%%%%%%
\preprint{smw-cfp1-09-07}
\title{Categorical Foundations for Physics - I: Program at a Glance}

%\title{Category Theory in Relativity} %EARLIER TITLE

\author{Sanjay M Wagh}

\altaffiliation{On absence from: Central India Research
Institute, Post Box 606, Laxminagar, Nagpur 440 022, India \\
E-mail: cirinag\_ngp@sancharnet.in or cirinag\_ngp@bsnl.in}

\affiliation{Astrophysics \& Cosmology Research Unit, School of
Mathematical Sciences, University of KwaZulu-Natal, Private Bag
X54001, Durban 4000
\\ E-mail: waghs@ukzn.ac.za}

%%%%%%%%%%%%%%% abstract below %%%%%%%%%%%%%%%%%%%%%%%%%%%%%%%%%%%
\begin{abstract}
Measures in the context of Category Theory lead to various
relations, even differential relations, of categories that are
independent of the mathematical structure forming objects of a
category. Such relations, which are independent of mathematical
structure that we may represent a physical body or a system of
reference with, are, precisely, demanded to be the Laws of Physics
by the General Principle of Relativity. This framework leads to a
theory for the physical entirety.
\end{abstract}
%%%%%%%%%%%%%%%%%%%%%%%%%%%%%%%%%%%%%%%%%%%%%%%%%%%%%%%%%%%%%%%%%

\date{September 5, 2007}
\maketitle

\section{Introduction}
We describe here a program of the categorical basis for Physics.
The overall motivation for this program arises out of following
considerations.

As Physics is our attempt to conceptually grasp the happenings
around us, and as Mathematics is a language for succinctly
expressing the associated conceptions, the most general
Mathematics would help us express the most general of such
conceptions. As a result, we would hopefully encompass the
entirety of physical phenomena within any such most general
description.

This point of principle was advocated along with the Universal
Theory of Relativity whose developmental stages can be found in
\cite{utr-foundations}. We also note that this universal
relativity is based on Einstein's general principle of relativity
that the Laws of Physics have the same mathematical form
irrespective of (the state of motion of) the system of reference,
any physical body. In difference with other attempts \cite{qg,
oziewicz} at providing a foundation for the physical entirety, we
seek to base universal relativity on the most general categorical
considerations for reasons arising out of this point of principle.

Then, in summary of the contents of \cite{utr-foundations}, what
is really needed is a way of dealing with and extracting the
needed ``information'' from the most general category because it
is the currently known most general mathematical structure. In
this context, an appropriate notion of (real-valued or not)
measures \cite{utr-foundations,cat-m} over a generic category is
the way.

\subsection{General Description of This Program}

The following part of Introduction specifically keeps in sight the
purpose of Mathematics for Physical Theories, rather than losing
it with details of mathematical considerations. In what follows,
we therefore discuss relevant conceptions without precise
mathematical details, which follow the end of this general
discussion of the involved issues.

To understand Nature, we ``associate'' a mathematical structure,
let it be any, with physical bodies, and study changes in that
mathematical structure to ``model'' changes in physical bodies.
When observations of Nature or results of concerned experiments
agree with the ``predictions'' of our model, we claim to
successfully ``explain'' the associated observable physical
phenomena.

Using different mathematical structures to represent physical
bodies, we also construct various such models, and compare them.
Then, a mathematical model explaining the largest body of
observational data with the least number of associated physical
conceptions is our fundamental understanding about Nature. We also
advance, mathematically easy, models that ``approximate'' the
fundamental model. The grand dream of physicists is of a
fundamental mathematical model, the Theory of Everything, which
explains {\em all\/} observations of Nature - the physical
entirety.

As is quite well known, Newtonian theories, even though they were
quite successful in explaining many observations of Nature, failed
to ``naturally'' explain observations related to electromagnetic
radiation, and various (quantum) experiments.

Einstein, in a remarkable insight, realized that Newtonian
theories were all based on the Special Principle of Relativity. By
additional principle of the constancy of the speed of light for
inertial systems of reference, he then formulated his Special
Relativity. Newtonian model is an approximation of this Special
Relativity. Successes of Special Relativity then convinced
\cite{einstein} Einstein that the ``strategy'' of ``Relativity''
for formulating model of the Physical World is truly an
appropriate one.

As a straightforward step, Einstein then proposed the General
Principle of Relativity, a statement of ``strategy'' that the Laws
of Physics be such that they have the same (mathematical) form
irrespective of the state of (motion of) the system of reference,
any physical body. With this strategy, Einstein, likewise with
Newton, aimed at a comprehensive theoretical description of the
physical entirety.

But, ``changes'' to mathematical structure representing physical
bodies also ``mean'' changes to reference physical bodies. The
(General or Universal) Principle of Relativity \cite{einstein,
utr-foundations} then requires those mathematical laws, to be the
Laws of Physics, which do not depend on even the mathematical
structure that we may associate with a physical body. This is then
``the'' meaning of General Covariance.

Now, any mathematical structure can be an ``object'' or an
``element'' of a collection of similar mathematical structures.
For example, a group, a ``structured'' set, can be an element of a
collection of {\em all\/} groups. Of course, we then have to avoid
set-theoretic paradoxes, like Russell's paradox, while making such
collections. This is always achievable by adopting a suitable
definition of a set such that a collection of all sets is not a
set, but what we ``name'' as a class. Furthermore, the collection
of all classes is also not a class, but what we ``name'' as a
conglomerate; and so on \cite{cat-b}. With these definitions,
Russell's paradox is then a harmless statement that a
``collection'' \footnote{Then, by a collection, we will always
mean a ``gathering'' or an ``accumulation'' of elements such that
set-theoretic paradoxes do not arise in its considerations.} of
all elements, gathered according to a defining property, does not
have the ``defining'' property of its elements.

A change of one mathematical structure to another of its
collection is called as an ``arrow'' connecting ``objects'' (of
that collection). An arrow then has a source (domain) and a target
(co-domain) as objects. Every arrow (relation) need not be a
function in the mathematical sense. An identity arrow of an object
is then an identity transformation of the associated mathematical
structure.

Mathematically, the collection of all such arrows forms a partial
binary algebra, and with, remarkably naturally arising, additional
compatibility properties, a Category \cite{cat-b}. It is a
mathematical structure completely specified \cite{cat-b} by only
the arrows. For a category, we also form sub-collections, called
as the {\em hom-collections}, of {\em all\/} the arrows from one
to another of its objects. This categorical structure is then
quite ``separate'' \footnote{That is why, in contrast to most of
the mathematical literature on this subject, we will call any such
structure simply as a category.} from the issues of the set
theory.

For example, a category of a group has only one object and group
elements as arrows. Every arrow of a group category is an
iso-arrow (isomorphism) of the only object of the group category,
for an inverse of every group element is a member of the group. A
category of a monoid also has only one object and monoid elements
as arrows. But, only the monoid identity is an iso-arrow of the
only object of this category, for the inverses of general monoid
elements are not members of the monoid.

The object-free structure of a category is then the fundamental
mathematical structure of {\em all\/} the mathematical structures,
as we can always form a collection of mathematical structures
similar to any chosen one and consider changes of one to others of
that collection as a category.

Different categories can now be ``related'' to each other by {\em
functors}, which are the partial binary algebra preserving maps
that also preserve identities and compositions of arrows. Functors
from one to another category always exist.

Notably, categorical can then be the mathematical structure of the
collection of all categories, when we form the category of all
categories \cite{lawvere}, for a category is a mathematical
structure with a functor changing it to another similar structure.

Now, we can ``reorganize'' arrows of any category in different
collections that can themselves be treated as (new) arrows, and by
ensuring that the conditions of the definition of a category are
satisfied, form another category. ``New'' categories from
``known'' categories are then obtainable.

Functors connecting two (fixed) categories can also be collected
to form a partial binary algebra, and a category, called as a
functor category, is also obtainable with ``natural
transformations'' or ``functor morphisms'' as arrows connecting
such functors.

Now, in the setting of the standard set theory, a set-theoretic
(standard or usual) measure is \cite{m-theory} a set-function that
forms a commutative monoid or an abelian group of addition over
(countable) collections of pairwise mutually disjoint sets.

Fundamental to set-theoretic measures are the additivity of
measures and the pairwise disjointedness of involved sets. It is
only when these hold that a measure is a unique association of a
set with an element of the (commutative monoid or) abelian group
of addition. We then use \cite{m-theory} ``other'' properties of
the monoid or group of addition, for example, topological ones, to
formulate our (general) notions of the continuity, the
differentiability, the metric etc.

For an abelian group of addition, the addition function, denoted
by $+$, is defined over the arrows with the same source and the
same target, and the composition of arrows of this group category
is left and right distributive over $+$, with the identity element
of the group of addition behaving as a ``zero arrow'' \cite{cat-b}
of this category with only one object. A category with this
additive structure is an {\em additive category}. An abelian group
and a commutative monoid of addition are its obvious examples.

In general, a category need not possess an ``additive structure''
over its collection of arrows, ``addition'' being a very specific
function on the collection of categorical arrows. Furthermore, no
natural notion of the complement of a categorical object is
available in a general category, even though sub-object of an
object corresponds to a subset of a set. Such difficulties had
prevented any definition of measures in the general categorical
context, till the work in \cite{cat-m}, whose strategy for
defining measures on a general category is what is described
below.

A general category, a collection of arrows, always has a
sub-collection of identity arrows of its partial binary algebra of
arrows. Then, we can always form sub-collections (families) of the
objects (identity arrows) of any category. Such a sub-collection
can also be empty, {\em ie}, we can have an empty family.
Intuitively, when we ``combine'' families to form a larger family,
we have the conception of ``addition'' implicit in this operation.

Then, if we ``reorganize'' the arrows of a category in conjunction
with the formation of the families of its objects, we obtain
\cite{cat-m} an additive structure of a commutative monoid over
the hom-collections of ``new'' arrows connecting the families.

Now, {\em co-product\/} is \cite{cat-b} a categorical
generalization of disjoint union of sets. In forming families of
objects of any category, we also generate \cite{cat-m} ``objects''
(families themselves) that are always the co-products of some
other objects (families) of their category. Then, a specific
additive category, to be called as a {\em pointed family category
of a given category}, can always be obtained from {\em any \/}
category by forming families of its arrows, and by forming
correspondingly ``new'' arrows for these families.

Next, the functor category of functors from the pointed family
category of a given category to any additive category has
\cite{cat-m} the additive structure of the latter category. Then,
the functor category of functors from the pointed family category
of a category to the category of a group (monoid) of addition of
real numbers has the additive structure of a group (monoid). It is
a unique association of an object of the former category with an
element of the concerned group (monoid). Such functors, henceforth
to be called as {\em measure functors}, are then (additive
category-based) categorical measures. Such categorical measures
are evidently definable for every category.

Notice here that the category of (group or monoid of) addition of
real numbers can also be replaced with any additive category.
Because functors map identities to identities, notice furthermore
that any measure functor assigns identity arrows of all the
objects of the pointed family category of a given category to only
one arrow, the identity arrow, which is also the only zero arrow,
of the additive category. Moreover, notice that functors preserve
compositions of arrows.

Measure functors that are ``naturally isomorphic'' to each other
form an equivalence class within the collection of all the measure
functors acting on a given category. All members of this
equivalence class provide an association of the ``same'' arrow of
the pointed family category of a category with the ``same'' arrow
of the additive category. Therefore, we need to consider a measure
functor, categorical measure, only modulo its equivalence class.

Now, a given category is ``embedded'' in its pointed family
category as a sub-category, and an ``inclusion functor'' describes
this embedding. (It is a one-one association of the arrows of the
involved categories.) Then, a composition of inclusion functor and
a measure functor is a functor, call it an {\em m-i functor\/}
(notice that an inclusion functor acts first, and then a measure
functor), from that given category to the additive category. Then,
such a composition of the involved functors, an m-i functor is
also an assignment of a (unique) element of additive category with
an object of that given category, it being independent of the
mathematical structure of the objects of that category.

An object of a category can itself be viewed as a category. Then,
there are ``internal'' measures as well as ``external'' measures
definable for it. Notice however that, as a collection of measures
definable for a category, the internal and external measures
belong to the same collection. [That these measures belong to the
same collection can be expected, for example, from the fact that
we can consider distance between two physical bodies, as well as
distance between ``parts'' internal to any physical body. These
are essentially the ``same'' mathematical notions, except that we
call one as ``external'' and the other as an ``internal'' distance
(measure). The same applies to other notions. Distinction between
internal and external measures is then a matter of our
convenience.]

Consider now an endo-functor, {\em ie}, a functor from a given
category to itself. Notice here that an endo-functor acts first,
then the inclusion functor and, in the end, the measure functor.
Endo-functors of any given category can then be classified as
``measure-preserving'' and ``measure non-preserving'' ones,
obviously, relative to the measure functor under considerations.
Then, the same endo-functor need not be preserving all the
measures of its category.

Now, topological structure of the additive (group or monoid)
category (of real numbers) permits definitions of continuity,
differentiability, metric, distance $\cdots$ Essentially,
categorical measures can be varied as per these notions, which are
available for every category. This is, quite naturally, as it is
with the usual or standard theory of measures.

An interplay of measure-preservation and non-preservation now
emerges as the mathematical basis underlying physical phenomena,
when we associate measures with various of our physical
conceptions. For example, consider a positive real-valued
``external'' measure to ``correspond'' to our notion of ``physical
distance'' between bodies. (We use topological properties of the
additive category while making any such correspondence.)
Mathematical law of measure non-preservation under the action of a
categorical endo-functor would then correspond to change in this
physical distance between corresponding physical bodies. Such a
law would, clearly, be a ``differential'' law, because the
topological structure of the additive category can be used to
write the action of an endo-functor in that manner.

Now, any such ``categorical differential structure'' too is
independent of the mathematical structure of the objects of a
category. This is \cite{catfp} then the precise ``mathematical
reason'' as to why the physical world is well describable using
differential equations.

Thus, we obtain categorical interrelationships that are, and only
such interrelationships can perhaps be, ``free'' of mathematical
structures forming a category. It may then be emphasized that
these interrelations arising out of the considerations of
categorical measures are {\em functorial\/} in character. This is
the primary reason for these categorical interrelationships to be
entirely independent of the mathematical structure of the objects
forming a category.

These categorical ``relations'' are then exactly what we needed to
mathematically implement Einstein's General Principle of
Relativity, for these are independent of the mathematical
structure that we associate with any physical body, and therefore,
free of any physical system of reference whatsoever. Clearly, such
``relations'' (of Universal Relativity) constitute then the most
general Laws about Nature \cite{catfp}. This is also the reason
why we can expect such laws to encompass the physical entirety.

Now, the notion of time is that of the periodic motion of a
``clock'' body relative to an ``observer'' body. Within the
present categorical framework, this notion of ``time'' is
obtained, using the notion of ``distance'' in this context, from
the ``periodic'' variations of the ``location'' of one
(categorical) object, the ``clock'' body, in relation to any other
(categorical) object, the ``observer'' body. [Notably, there does
not exist any other notion of ``time'' within this general
categorical context.] Categorical ``velocity'', ``acceleration'',
etc.\ then use such a notion of time.

Of categorical measures is then also the mathematical way using
which we can \cite{catfp} now ``understand'' how one model of
Nature ``approximates or not'' another model, because measures
provide mathematically precise sense of ``closeness'' of models.

With this introduction, we now turn to mathematical considerations
of this program for the categorical basis of the physical
entirety.

In Section \ref{catmeasure}, we consider measures in the
categorical context. In Section \ref{cffp}, we first work out an
explicit example of the physical distance measure to establish the
use of present framework for physical considerations. We also
briefly mention the general procedure for obtaining the Laws of
Physics from the considerations of categorical measures. But, the
general procedure of using categorical measures for assigning
properties to physical bodies is quite involved. It therefore
deserves separate presentation \cite{catfp}. We end this article
with some remarks in Section \ref{conclude}.

\section{Categorical Measures} \label{catmeasure}
A fundamental property of standard or usual or set-theoretic
measures is their (countable) additivity. Categorically, only the
``additivity'' is defined as follows:

\begin{defin} A \underline{monoid} (g\underline{rou}p)
\underline{additivit}y structure on any p\underline{ointed}
category $\cat{A}$ is a function $+$ that
associates with each pair $A\dmorphr{g}{h}B$ of %(parallel)
$\cat{A}$-arrows with common source (or domain) $A$ and common
target (or co-domain) $B$, another $\cat{A}$-arrow, denoted by
$g+h:A\to B$ or by $\displaystyle{A\;\morph{g+h}\;B}$, such that

\begin{description} \item{(M1)} for each pair $(A,B)$ of
$\cat{A}$-objects, the function $+$ induces on the corresponding
hom-collection $[A,B]$ a commutative structure of a monoid (an
abelian group) of addition,
\item{(M2)} the composition of arrows in $\cat{A}$ is left and right
distributive over $+$, {\em ie}, whenever $C\morph{k} A
\dmorphr{g}{h} B \morph{f} D$ are $\cat{A}$-arrows, then $f\circ
(g+h) = (f\circ g) + (f\circ h)$ and $(g+h)\circ k = (g\circ k) +
(h\circ k)$. That is, the composition functions $[B,C]\times [A,B]
\to [A,C]$ are bilinear.
\item{(M3)} The zero arrows of $\cat{A}$ act as monoid (group) identities
with respect to $+$, {\em ie}, for each $\cat{A}$-arrow $f$,
$0+f=f+0=f$. That is, the identity elements of the monoids
(abelian groups) behave as zero arrows whenever the compositions
are defined.
\end{description} \end{defin}

The category $\cat{G}$ of the commutative group as well as the
category $\cat{M}$ of a commutative monoid are obvious examples of
categories having the above additive structure(s). In general, we
will denote  by $\cat{A}$ any category possessing a (monoid or
group) additive structure.

To ``construct''  \cite{cat-m} an appropriate category
$\cat{F}amily(\cat{C})$ of the families of arrows of a generic
category $\cat{C}$, let $A_I$ be the family $(A_i)_{i\in I}$ of
objects of $\cat{C}$, indexed by some collection $I$. With
families $A_I$ as objects, an arrow from $A_I$ to $B_J$ is then a
pair $(f,\mathfrak{f})$ with $f: I \to J$ as a map of collections
and $\mathfrak{f}$ as a family $\mathfrak{f}: \left( A_i
\morphr{\mathfrak{f}_i} B_{j(i)}\right)_{i\in I}$ of arrows in
$\cat{C}$. For arrows $(f,\mathfrak{f}):A_I \to B_J$ and
$(g,\mathfrak{g}):B_J \to C_K$ in the category
$\cat{F}amily(\cat{C})$, the composition $(g,\mathfrak{g})\circ
(f,\mathfrak{f})$ is defined as the arrow $(h,\mathfrak{h}):A_I
\to C_K$ such that $h=gf$ and $\mathfrak{h}_i =
\mathfrak{g}_{f(i)}\mathfrak{f}_i$. An identity arrow for $A_I$ in
$\cat{F}amily(\cat{C})$ is the identity map $\id_I:I\to I$ and the
family, $\{\id_{A_i} \}_{i\in I}$, of identity arrows for
$\cat{C}$-objects $A_i$, $i\in I$. By this construction, any
category $\cat{C}$ is always {\em fully embedded\/} in the
category $\cat{F}amily(\cat{C})$, its family category, as a
sub-category.

Every object $A_I$ of the category $\cat{F}amily(\cat{C})$ is a
co-product of its constituent objects $\{A_i\}_{i\in I}$ (viewed
as one-member families) with $\mathbb{I}_i=(i, 1_{A_i}):A_i \to
(A_i)_{i\in I}$ being co-product injections. An initial object of
$\cat{F}amily(\cat{C})$ is an ``empty'' family, while its terminal
object is the ``singleton'' family, and these two are distinct
objects of category $\cat{F}amily(\cat{C})$. Hence,
$\cat{F}amily(\cat{C})$ is not a pointed category. It therefore
does not have additive structure defined above.

A pointed category $p\cat{F}amily(\cat{C})$, freely constructible
from the category $\cat{F}amily(\cat{C})$, has objects as pairs
$(A_I, A_i)$ where $A_I$ is an object of $\cat{F}amily(\cat{C})$
indexed by $I$ with (fixed) $A_i\in A_I$ being a ``base'' object
of this pair. As its arrows, this category has the arrows of
$\cat{F}amily(\cat{C})$ of the form $(f,\mathfrak{f}) :A_I\to B_J$
such that we also have $(f,\mathfrak{f})A_i =B_j$ with $A_i$ and
$B_j$ being taken as one-member families. That is, we have
$f(i)=j$, and there always exists an arrow $\mathfrak{f}_i:A_i\to
B_j$ in the collection $\mathfrak{f}$. We say that any such arrow
``preserves'' the bases of the families. Its hom-collection,
$\left[ \left(A_I,A_i\right), \left(B_J,B_j\right) \right]$, is a
collection of all such arrows in $\cat{F}amily(\cat{C})$. An
identity arrow $\left(\mathrm{id}_I, \{\mathrm{id}_{A_i} \}_{i\in
I}\right): \left(A_I,A_i\right) \to \left(A_I,A_i\right)$ also
exists in such collections with obviously $\left(\mathrm{id}_I,
\{\mathrm{id}_{A_i} \}_{i\in I}\right) A_i=A_i$. An object
$(\{\emptyset\}, \emptyset)$ is a zero object of
$p\cat{F}amily(\cat{C})$, that was mentioned earlier as the
pointed family category of a category $\cat{C}$. The zero arrows
of $p\cat{F}amily(\cat{C})$ are the arrows with empty domain, and
these are its only zero arrows. The category
$p\cat{F}amily(\cat{C})$ therefore has the additive structure
defined before. Every object $(A_I,A_i)$ of the category
$p\cat{F}amily(\cat{C})$ is also a co-product of its constituent
objects $(\{A_i\},A_i)_{i\in I}$.

Also, the category $\cat{C}$ is embedded in the category
$p\cat{F}amily(\cat{C})$, and the corresponding inclusion functor
$\funct{I}: \cat{C} \to p\cat{F}amily(\cat{C})$ is given by the
one-one association of any arrow $A\morphr{g}B$ of category
$\cat{C}$ with an arrow $(\{A\},A) \morphr{(f,\{g\})} (\{B\},B)$
with $(f,\{g\})A =B$ of the category $p\cat{F}amily(\cat{C})$.
However, the inclusion of a sub-category $\cat{C}$ in its pointed
family category or its pointed free co-product completion category
$p\cat{F}amily(\cat{C})$ is not, always, a {\em full\/} embedding,
even though the inclusion of a sub-category $\cat{C}$ in its
family category $\cat{F}amily(\cat{C})$ is always a full
embedding.

Now, if a category $\cat{A}$ is additive, then, for any category
$\cat{B}$, the functor category, $\funcat{B}{A}$, inherits the
additive structure of $\cat{A}$. Thus, a functor category
$\cat{G}^{p\cat{F}amily(\cat{C})}$ has the group additive
structure, while the functor category
$\cat{M}^{p\cat{F}amily(\cat{C})}$ has a monoid additive
structure.

The additive structure of $\cat{A}^{p\cat{F}amily(\cat{C})}$ is
precisely that of the natural transformations of functors from
$p\cat{F}amily(\cat{C})$ to $\cat{A}$. The additive structure of
$\cat{A}^{p\cat{F}amily(\cat{C})}$ is then the categorical
analogue of the ``countable additivity'' of usual measures.
Therefore, $\cat{A}$-based measures, also called $\cat{A}$-based
measure functors or simply as categorical measures, on an
arbitrary category $\cat{C}$ are now defined to be the objects of
the functor category $\cat{A}^{p\cat{F}amily(\cat{C})}$.

Notably, what we require for the definition of categorical
measures are the ``additivity'' of category $\cat{A}$ relative to
which measures are defined, and the ``co-product completion
character'' of ``additive'' category $p\cat{F}amily(\cat{C})$ that
is ``freely constructible'' from any arbitrary category $\cat{C}$.
Categorical analogue of the ``countable additivity'' of these
categorical measures trivially follows because such a functor
category is ensured to have an appropriate additive structure.

In general, a symbol $\funct{M}$ will be used to denote an
$\cat{A}$-based categorical measure or a measure functor
$\funct{M}: p\cat{F}amily(\cat{C}) \to \cat{A}$. All functors,
whenever used, will be considered modulo all those functors that
are {\em naturally isomorphic\/} to each other.

Now, composition of a measure functor $p\cat{F}amily(\cat{C})
\morphr{\funct{M}} \cat{A}$ and an inclusion functor $\cat{C}
\morphr{\funct{I}} p\cat{F}amily(\cat{C})$ is a functor $\cat{C}
\morphr{\funct{M\circ I}} \cat{A}$. Because an inclusion functor
associates with an arrow of category $\cat{C}$ a {\em unique\/}
arrow of category $p\cat{F}amily(\cat{C})$, and a measure functor
associates with an arrow of $p\cat{F}amily(\cat{C})$ a unique
arrow of the additive category $\cat{A}$, the functor
$\funct{M\circ I}$, to be called as an {\em m-i functor},
associates a unique arrow of category $\cat{A}$ with an arrow of
category $\cat{C}$. An m-i functor is then an association of a
unique element of the additive monoid or group with an arrow of
the category $\cat{C}$. [When the arrow under consideration is an
identity arrow, also called as a unit of the partial binary
algebra of arrows, in the category $p\cat{F}amily(\cat{C})$, a
measure functor associates it with an identity arrow of $\cat{A}$,
with this association being clearly independent of the
mathematical structure of the object, which only ``labels'' that
identity arrow.] Then, we will also call an m-i functor as an
$\cat{A}$-based measure or categorical measure on the objects of
category $\cat{C}$. An m-i functor $\funct{M\circ I}:\cat{C} \to
\cat{A}$ is a partial binary algebra preserving, an
identity-preserving and compositions-preserving {\em function\/}
from the collection, $\mathcal{C}(\cat{C})$, of all the arrows of
a category $\cat{C}$ to the collection, $\mathcal{C}(\cat{A})$, of
all the arrows of an additive category $\cat{A}$.

In terms of the standard ways of measure theory, we have then
effectively {\em isolated}, as corresponding measures, a
collection of countably additive functions $\funct{M\circ
I}:\mathcal{C}(\cat{C}) \to \mathcal{C}(\cat{A})$, which are
partial binary algebra preserving, identity preserving and
compositions preserving.

Families of the units of the partial binary algebra of the
category $\cat{C}$, families of its objects, do not however
correspond to Borel sets \cite{m-theory}. We can, from an object
$A$ in $\cat{C}$, form a family $A_I=\{A,A,A, \cdots\}$ that is
non-Borel for the usual ways of measure theory.

The usual Borel structure however exists with the hom-collections
$[(A_I,A_i), (B_J, B_j)]$ of the category
$p\cat{F}amily(\cat{C})$, as we are allowed to form a Borel
structure for {\em every\/} such hom-collection because of the
very definition \cite{cat-b} of a category. The construction of
the category $p\cat{F}amily(\cat{C})$ serves here to ``ensure''
therefore only the ``consistency of structures'' (generally
unavailable for any arbitrary category) of hom-collections of the
category $p\cat{F}amily(\cat{C})$ and that of an additive category
$\cat{A}$. We are therefore allowed the use of the standard
measure theory, limited to aforementioned considerations of the
hom-collections, of course. Categorical measures, then, provide
the overall consistency of categorical structures containing all
the hom-collections of the involved categories.

Of particular interest are the hom-collections $[(\{A\},A),
(\{A\}, A)]$ for an arbitrary object $A$ of the category
$\cat{C}$. In the case that this hom-collection has only a single
arrow, an identity arrow, any categorical measure (functor) must
map it to the additive identity of the additive category. For
categorical measures, such are then the situations of categorical
measure zero.

In general, each of the hom-collections $[(A_I,A_i), (B_J, B_j)]$
of $p\cat{F}amily(\cat{C})$ is therefore a Borel space with
associated Borel structure. It is of course the ``same'' measure
that gets ``defined'' here for all the hom-collections of the
category $p\cat{F}amily(\cat{C})$.

Then, under conditions of the Lebesgue-Radon-Nikodym (LRN) Theorem
\cite{m-theory}, for each of these hom-collections, there exists a
unique finite valued measurable function, $f={d\ell}/{dt}$, the
LRN-derivative, where the $\sigma$-finite measure $\ell$ is
absolutely continuous with respect to measure $t$, and all
properties of the ``differential'' hold modulo a set of
$t$-measure zero, {\em ie}, $t$-almost everywhere or $t$-a.e.

Any categorical measure ``carries'' this ``differential
structure'' consistently to the category $p\cat{F}amily(\cat{C})$
and, thence, to the category $\cat{C}$ by way of the composition
$\funct{M\circ I}$.

Because an object of a category can itself be looked upon as a
category in its own right, we have ``categorical measures'' or
$\cat{A}$-based measures that are ``internal'' to an object, and
``measures'' that are ``external'' to it. Internal and external
measures of any categorical object belong however to the same
collection of $\cat{A}$-based categorical measures definable for
any category. As was remarked earlier, this feature is not
surprising because ``externally'' usable notions can also be used
``internally'' to a (physical as well as mathematical) object.
Nevertheless, internal measures ``characterize'' categorical
objects within this most general categorical framework.

A group of addition of real numbers, whose set is denoted by
$\mathcal{R}$, is a topological group $(\mathcal{R},+)$ with
respect to the usual metric topology. A monoid of addition of real
numbers, $\mathcal{R}_+$ being the set of strictly positive (or
strictly negative) real numbers, is a topological monoid
$(\mathcal{R}_+,+)$ with respect to the usual metric topology.
Both, $(\mathcal{R},+)$ and $(\mathcal{R}_+,+)$ are locally
compact. $\cat{R}^+$ denotes the group category, and $\cat{R}^+_+$
the monoid category, of addition of real numbers.

Then, the collection $\mathcal{C}(\cat{A})$ of all the arrows of
an additive category has the structure of a locally compact
topological monoid, for example, of $(\mathcal{R}_+,+)$ under the
usual metric topology. Due to its construction, the collection
$\mathcal{C}(p\cat{F}amily(\cat{C}))$ of the arrows of the
category $p\cat{F}amily(\cat{C})$ has the structure of the product
of locally compact topological monoids, and has the associated
product topology. Therefore, categorical measures $\funct{M}:
\mathcal{C}(p\cat{F}amily (\cat{C})) \to \mathcal{C}(\cat{A})$ are
the partial binary algebra preserving, identity preserving, and
compositions-preserving {\em continuous\/} functions from
$\mathcal{C}(p\cat{F}amily(\cat{C}))$ to $\mathcal{C}(\cat{A})$.

\section{Categorical Foundations for Physics} \label{cffp}
Now, we may ``visualize'' a mathematical structure, let it be any,
representing physical bodies as being an object of a category, and
the changes or the transformations of physical bodies as the
arrows of that category. Then, entirely independently of the
mathematical structure that is chosen to represent physical
bodies, categorical measures of categorical objects represent
various of the physical properties of physical bodies within this
categorical framework. Consequently, categorical measures are then
``fundamental'' to formation of any of our physical conceptions.

The characterization of categorical objects by internal measures,
and the overall framework of category theory, then provide
``relations'' that are, in a definite sense, ``free'' of the
mathematical structure that we choose to represent physical bodies
with. This is the {\em categorical general covariance}. In
complete conformity with the General Principle of Relativity, such
relations then constitute the most general Laws about Nature. We
can expect such categorical laws to encompass the physical
entirety. In particular, such laws turn out to be differential
laws. This is then the precise mathematical reason as to why the
physical world is well describable using differential equations.

Aforementioned characteristics of (categorical) objects are the
``observable characteristics'' of physical bodies within this most
general mathematical framework. Changes in these characteristics
are then the changes to physical bodies. The essential aim of a
Program of Categorical Foundations for Physics is then that of
obtaining definite mathematical laws for the changes (modulo their
``equivalence'' classes) of aforementioned characteristics
(measures) of categorical objects.

Now, the categorical procedure for obtaining
``object-independent'' relations using categorical measures is
required to be such as to be applicable to all the relations so
obtainable. This, in fact, is the pivotal issue, which underlies
the categorical general covariance.

To fix ideas, consider therefore any categorical measure
$\funct{M}: \mathcal{C} (p\cat{F}amily (\cat{C})) \to
\cat{R}^+_+$. Its additivity property allows the corresponding
``metric'' structure to be constructed. We now ``actually
construct'' a metric structure that is the one of (physical)
distance separating the objects.

To this end, call the  families $A^{SF}_I=\{A,A,\cdots\}$ as {\em
self-families\/} of the object $A$ of $\cat{C}$. Consider now a
categorical measure $\funct{D}$ that, for every object $A$ of the
category $\cat{C}$, maps {\em every\/} of the hom-collections
$[(A^{SF}_I,A),(A^{SF}_J,A)]$ of $p\cat{F}amily(\cat{C})$ to the
``same'' element, {\em additive identity}, of the additive
category $\cat{R}^+_+$. Since the hom-collection $[(\{A\},A),
(\{A\},A)]$ also gets mapped to zero of real numbers, we call this
as the property of vanishing of the ``self-distance'' of objects.

Moreover, for this {\em physical distance measure}, $\funct{D}$,
and for non-identical (which could be isomorphic) objects $A$ and
$B$ of $\cat{C}$, we require that all the hom-collections
$[(A^{SF}_I,A), (B^{SF}_J,B)]$ are mapped by it to the ``same''
non-identity element, say $a$, of the additive category
$\cat{R}^+_+$. Consistency with the vanishing self-distance
property is then self-evident. The element $a$ of $\cat{R}^+_+$
then defines metrical distance, $d(A,B)$, between objects $A$ and
$B$ of category $\cat{C}$, with metrical properties of $d(A,B)$
following from the additivity of the physical distance measure
$\funct{D}$. We associate the metric structure of measure
$\funct{D}$ with that of the {\em physical distance\/} separating
physical bodies represented by the objects of category $\cat{C}$.

Physically, the aforementioned has the significance that the
``distance'' of a physical body from itself would then be
vanishing always, and furthermore making any family out of that
physical body would be of no physical relevance to this vanishing
self-distance.

Such a physical distance measure, a functor, evidently exists, and
the metric structure corresponding to its additivity defines the
physical distance separating physical bodies.

The aforementioned is an instance of a general procedure for
defining ``characterizing'' properties for physical bodies.
Evidently, it involves ``classifying'' categorical measures
according to their actions on the objects of the category
$p\cat{F}amily(\cat{C})$. It will be discussed in \cite{catfp}.

Now, changes to categorical measures $\funct{M}: \mathcal{C}
(p\cat{F}amily (\cat{C})) \to \cat{R}^+_+$ can occur when an
endo-functor of the category $\cat{C}$, a functor $\funct{E}:
\cat{C} \to \cat{C}$, ``changes'' the assignments of its arrows
(with identity arrows). An endo-functor is 1-1 and onto on the
collection $\mathcal{C}(\cat{C})$ of arrows of the category
$\cat{C}$.

Let the physical distance measure possess the assignments:
$d(A,B)=a$, $d(B,C)=b$, $d(C,A)=c$.  Endo-functor can, for
example, contain a ``cycle'' of the form $\funct{E}A= B$,
$\funct{E}B= C$, $\funct{E}C=A$, while mapping all the other
objects of $\cat{C}$ to themselves. Then, under the action of
endo-functor $\funct{E}$, we have $d(\funct{E}A,\funct{E}B)=b$,
$d(\funct{E}B,\funct{E}C)=c$, $d(\funct{E}C,\funct{E}A)=a$, {\em
ie}, distance between $A$ and $B$ thus changes from being $a$ to
being $b$ etc. Distances of $A$, $B$, $C$ from all other objects
also change, without any changes to mutual distances of all other
objects. Now, if $\funct{F}$ is another endo-functor that restores
``original'' distances, and such a functor obviously exists, then
we have an ``instance'' of periodic motion under the action of
composition of endo-functors $\funct{F\circ E}$.

A general endo-functor then causes changes to
$p\cat{F}amily(\cat{C})$, and therefore to measures from
$p\cat{F}amily(\cat{C})$ to $\cat{R}^+_+$. Periodic behavior of a
general endo-functor can then result in a periodic change in the
``physical distance'' of one from other objects. This describes a
periodic ``motion'' of an object. By the continuity of the
distance measure, these are continuous changes. The period of such
a motion then provides us the notion of {\em time}, importantly
{\em relative to that periodic motion}. Then, time is also a
suitable categorical measure within this framework.

An ``absolute zero'' of time is thus the situation when there is
no motion ``whatsoever'' of any of the objects of category
$\cat{C}$. It can evidently be ``reached'' more than once, {\em
ie}, when some motions of objects take place, then all the motions
of {\em all\/} bodies stop, and then some motions occur; again and
again. [There is then no ``origin'' of the ``Universe'' of
physical bodies within this framework, just as it was the
situation with Newton's theoretical framework.]

Alternatively, we may consider Borel automorphisms \cite{ergodic}
of the (Borel) space $\mathcal{C} (p\cat{F}amily(\cat{C}))$. A
Borel automorphism is \cite{ergodic} then ``periodic'' if every
point of the Borel space is periodic, but the period may differ
from point to point of the Borel space. [Every point of the space
$\mathcal{C} (p\cat{F}amily(\cat{C}))$ is periodic because every
``distance'' related to an object undergoing periodic motion
changes periodically, and it is ``object-symmetric''.] Then, we
have the following.

The notion of time in Physics is that of the periodic motion of a
``clock'' body relative to an ``observer'' body. Within the
present categorical framework, this notion of ``time'' is
obtained, using the notion of ``distance'' in this context, from
the ``periodic'' variations of the ``location'' of one
(categorical) object, the ``clock'' body, in relation to any other
(categorical) object, the ``observer'' body. [{\em Notably, there
does not exist any other notion of ``time'' within this general
categorical framework using measures.}] Categorical ``velocity'',
``acceleration'', etc.\ then use such a notion of time. These
mathematical details will be communicated separately \cite{catfp}.

In the above, we have provided an example of the way in which
categorical measures can be employed to describe the physical
phenomena - physical changes. Of course, we considered only one
type of a measure - the physical distance measure - for this
example.

Under the action of an endo-functor of a category $\cat{C}$, other
types of measures may also change. These changes then provide
descriptions of other physical phenomena associated with those
measures. An interplay of measure-preservation and
non-preservation now emerges as the mathematical basis underlying
physical phenomena. The aforementioned formalism uses the metric
structure associated with additivity of categorical measures, and
that provides the basis for relationally obtaining the Laws of
Physics. There does not appear to be any another way in which Laws
of Physics can be obtained within the proposed categorical
framework of universal relativity. Therefore, this categorical
description of the physical universe is entirely based on ``mutual
relationships'' of objects.

Now, fundamental constants arise when we consider relations of
physical bodies with each other. For example, Newton's constant of
gravitation arises when we ``describe'' the fall of a body to the
Earth as being due to the gravitational ``force'' of the Earth on
it. When we express this equationally, the proportionality factor
is this constant. We call it as a ``fundamental'' constant,
because it arises in considerations of the ``source'' property of
the force of gravitation. (The concept of Force requires a
physical body to have the property to generate it - the source
property. For gravity, it is the gravitational mass. Such a
property is an assumption, which cannot be explained by a theory
using the concept of Force.) Then, fundamental constants arise,
precisely, in considerations of various mutual ``relations'' of
physical bodies, and relate to their ability to ``act'' on each
other.

Thus, if a fundamental constant is undecidable in a theory, it is
then implied that ``some relationships'' of physical bodies of the
observable world are not within its explanatory powers. An
immediate conclusion is then that the theory in question cannot
describe physical entirety.

Any theory is a description of the physical world, in the language
of Mathematics as a logical-deductive system of conceptions. Then,
if a theory has an undecidable fundamental constant in it, it is
implied that its conceptions and its (mathematical) language,
both, are inadequate  to describe ``some relationships'' of
physical bodies of the observable world.

Within the proposed categorical framework, fundamental constants
can only arise from the relations of (categorical) objects. This
is an indication that the proposed categorical framework (using
measures) of universal relativity is that of a Theory of
Everything.

The program of categorical foundations for Physics then consists
of using provided (and therefrom derived) notions to obtain
\cite{catfp} the most general laws of Physics.

\section{Concluding Remarks} \label{conclude}
To conclude, we have outlined here a specific categorical program
for the foundations of physical entirety. It is based on the
concept of categorical measures, which are definable for an
arbitrary category. Evidently, it is entirely {\em independent\/}
of the mathematical nature of categorical objects, a statement
which is the categorical equivalent of the usual general
covariance. This approach, that can be rightfully called as the
Universal Relativity, is then in conformity with Einstein's
General Principle of Relativity. It is the currently known most
general mathematical framework for Physics.

One may now ``conjecture'' that all the categorical velocities
(obtainable from categorical measures) form, in general, a
groupoid category \cite{cat-b}. Provided that this ``conjecture''
holds, the formulation in \cite{oziewicz} can be one (categorical)
representation of universal relativity. Of course, categorical
measures, which are the fundamental mathematical notion than those
derived ones such as a velocity, then provide (physical)
properties of the objects of a groupoid category, which may be
said to ``describe'' the ``kinematical'' part of universal
relativity. The ``internal'' measures will, however, be outside
the scope of this kinematical representation of universal
relativity.

Now, an interesting laser interferometry experiment recently
showed \cite{braxmaier} a non-detection of a frequency shift, to
be specific, $\delta\nu/\nu \approx (4.8\pm5.3)\times10^{-12}$
over an observation period of $\sim 200$ days. This is then also
the fractional change, $\delta c/c$, in the speed of light, $c$.
Then, the non-detection of the frequency-shift in the experiment
of \cite{braxmaier} indicates that the speed of light (in vacuum)
is a {\em fundamental constant\/} that is independent of the
system of reference.

Importantly, the constancy of the speed of light irrespective of
the system of reference is an immediate consequence of the
groupoid category formulation \cite{oziewicz}.

Discussed here are therefore some ``features'' of a Theory of
Everything then.

\acknowledgments It is my pleasure to acknowledge Sunil Maharaj,
Partha Ghosh, Dharms Baboolal, Sudan Hansraj and other colleagues
in Durban for useful discussions, constant encouragement and
support. In particular, I am indebted to Sunil Maharaj and Partha
Ghosh for their insights during discussions, and to Zbigniew
Oziewicz for insightful correspondence. I am also grateful to
arxiv-moderation for suggestions, in particular of providing
mathematical background of universal relativity together with the
general description of the proposed program, that led to better
presentation.

\goodbreak


\begin{thebibliography}{99}

\bibitem{utr-foundations} See for the overall development of concerned
ideas the following and references therein:
\\ Wagh S M and Wagh A H (2006) {\it The Gravity and The Quantum:
A Bohr-inspired Synthesis}, available as ({\bf physics/0608027})
\\ Wagh S M (2005) {\em Universal Relativity and Its Mathematical
Requirements}, Talk delivered at the South African Mathematical
Society's (SAMS) 48th Annual Meeting, Grahamstown, October 31 -
November 2, 2005 ({\bf physics/0602038}) \goodbreak Wagh S M
(2005) {\em Progress with a Universal Theory of Relativity}, Talk
delivered at the South African Relativity Society's (SARS)
Einstein Centennial Meeting, Durban, September 25-26, 2005  ({\bf
physics/0602032}) \\ Wagh S M  (2005) {\em Foundations of a
Universal Theory of Relativity\/} ({\bf physics/0505063})

\bibitem{qg} D\"{o}ring A and Isham C J (2007) {\em A Topos Foundation
for Theories of Physics - I, II, III, IV}, available as ({\bf
quant-ph/0703060}), ({\bf quant-ph/0703062}), ({\bf
quant-ph/0703064}), ({\bf quant-ph/0703066})

\bibitem{oziewicz} Oziewicz, Z (2007) {\em Int. J. Geom. Methods
in Mod. Phys.}, ({\bf 4}), {\bf 4}, 1-11, ({\bf math/0608770}) \\
Oziewicz, Z (2006) {\em The Lorentz Boost is not unique}, (Fifth
Workshop: Applied Category Theory, Graph-Operad-Logic, Merida, May
2006) ({\bf math-ph/0608062})

\bibitem{cat-m} Wagh S M (2006) {\em Measures in the Categorical
Context}, preprint: smw-cat-m-02-06, Preprint circulation date:
September 2, 2006.

\bibitem{einstein} Einstein A (1970) in {\it Albert Einstein:
Philosopher Scientist} (Ed. P A Schlipp, Open Court Publishing
Company - The Library of Living Philosophers, Vol VII, La Salle)
\\ Einstein A (1968) {\it Relativity: The Special and the
General Theory\/} (Methuen \& Co. Ltd, London) (Appendix V:
Relativity and the Problem of Space.) \\ See relevant topics also
from, Pais A (1982) {\it Subtle is the Lord ... The science and
the life of Albert Einstein} (Clarendon Press, Oxford)

\bibitem{cat-b}
Ad$\acute{a}$mek J, Herrlich H and Strecker G E (2004) {\it
Abstract and Concrete Categories - Joy of Cats\/} and references
therein. Online edition available as: ({\bf
katmat.math.uni-bremen.de/acc}) \\ Barr M and Wells C (2002) {\em
Toposes, Triples and Theories}, (Version 1.1, 7 November 2002).
Online edition available as: ({\bf
www.cwru.edu/artsci/math/wells/pub/ttt.html}) \\ Mac Lane S (1998)
{\em Categories for the Working Mathematician}, (2nd Edition,
Springer-Verlag, New York) and references therein \\ MacLane S and
Moerdijk I (1992) {\em Sheaves in Geometry and Logic - A First
Introduction to Topos Theory}, (Springer-Verlag, New York)

\bibitem{lawvere} Lawvere F W (1966) {\em The category of categories as a
foundation for mathematics}, (La Jolla Conference on Categorical
Algebra, Springer-Verlag) pp. 1 - 20

\bibitem{m-theory} Halmos P R (1978) {\em Measure Theory} (Springer
International Student Edition, Narosa Publishing House, New Delhi)
and references therein. \\ Higgins P J (1974) {\em Introduction to
Topological Groups}, (London Mathematical Society Lecture Note
Series 15, Cambridge University Press, Cambridge) and references
therein. \\ Rohlin V A (1966) {\it American Mathematical Society
Transactions}, {\bf (2) 49}, 171 \\ Rohlin V A (1962) {\it
American Mathematical Society Transactions} {\bf (1) 10}, 1

\bibitem{catfp} Wagh S M (2007) Wagh S M (2007) {\em Categorical Foundations for
Physics - II: Laws of Physics}, in preparation.
\\ Wagh S M (2007) {\em Categorical Foundations for Physics - III:
Approximate Models}, in preparation. \\ Wagh S M (2007) {\em
Categorical Foundations for Physics - IV: Some Experimental
Tests}, in preparation. \\ Wagh S M (2007) {\em General Relaivity
and Category Theory}, to be submitted.

\bibitem{ergodic} Nadkarni M G (1995) {\em Basic Ergodic Theory},
(Texts and Readings in Mathematics No. 6, Hindustan Book Agency,
New Delhi), and references therein. \\ For duality involving
Borel spaces and boolean $\sigma$-frames, see Baoolal, D and
Ghosh, Partha Pratim (2007) {\em A Duality involving Borel
Spaces}, {\it J. of Algebraic Programming}, to be published.

\bibitem{braxmaier} Braxmaier C, M\"{u}ller H, Prandl O, Mlynek J
and Peters A {\it Phys. Rev. Lett.}, {\bf 88}, 10401 (2002)
\end{thebibliography}
\end{document}